# DDCCNet: Physics-enhanced Multitask Neural Networks for Data-driven Coupled-cluster

P. D. Varuna S. Pathirage, Konstantinos D. Vogiatzis*

*Department of Chemistry, University of Tennessee, 37996 Knoxville, Tennessee, United States*

*Corresponding Author: kvogiatz@utk.edu

**Abstract**

We present the data-driven coupled-cluster deep network (DDCCNet), a family of multitask, physics-enhanced deep learning architectures designed to predict coupled-cluster singles and doubles (CCSD) amplitudes and correlation energies from lower-level electronic structure methods. The three DDCCNet variants (termed as v1, v2, and v3) progressively incorporate architectural refinements ranging from parallel subnetworks for $t_1$ and $t_2$ amplitudes to feature-partitioned blocks and physics-enhanced intermediate prediction layers that are structured in accordance with coupled-cluster equations to enhance physical consistency and multitask learning efficiency. These models jointly learn correlated amplitude patterns while embedding symmetry and orbital-level interactions directly into the network structure. Applied to methanol conformers, $CO_2$ clusters, and small organic molecules, DDCCNet_v2 delivered the most accurate and transferable performance, achieving chemically precise correlation energies across diverse molecular systems. Collectively, DDCCNet establishes a scalable, physically grounded framework that unifies machine learning and ab initio theory for efficient, data-driven electronic structure prediction.

**Introduction**

Quantum chemistry describes the structure, properties, and reactivity of molecules by solving, either exactly or approximately, the many-electron Schrödinger equation. At the heart of this problem lies electron correlation,[1] the intrinsically coupled motion of electrons arising from their mutual Coulomb repulsion, which is not captured by a single-determinant mean-field description such as Hartree-Fock (HF) theory. The formally exact solution within a given one-particle basis is provided by full configuration interaction (FCI), which includes all possible electronic excitations and thus yields the exact correlated ground state.[2] However, the factorial scaling of FCI with system size renders it infeasible beyond the smallest molecules, motivating the development of hierarchical approximations that capture different components of correlation. From a physical perspective, electron correlation can be separated into strong correlation, arising from near-degeneracies and multireference character, and dynamic correlation, associated with short-range electron-electron repulsion around a dominant reference. Within this framework, coupled-cluster (CC) theory provides a formally exact description of electron correlation when all excitation ranks are included, in which case it becomes equivalent to FCI.[3] However, the steep computational scaling of the full CC expansion makes such it computationally prohibitive, requiring systematic truncations of the excitation manifold. Among these, the coupled-cluster singles-and-doubles with perturbative triples, or CCSD(T),[4] is widely regarded as the method of choice for systems dominated by dynamic correlation and is often considered the benchmark of quantum chemistry because of its exceptional balance between accuracy and computational cost. Nevertheless, accurate CCSD(T) calculations are limited to small- and medium-size molecules since the computational time, memory footprint, and disc storage requirements increase steeply as the number of correlated electrons increases. In addition, coupled-cluster methods rely on iterative solvers that involve computationally demanding, repetitive tensor contractions.

Artificial intelligence (AI) and machine learning (ML) have recently proposed as promising strategies to overcome these limitations.[5, 6] While machine-learning interatomic potentials (MLIPs)[7] aim to learn effective interactions between atoms in their local environments, ML frameworks that target correlated electronic structure methods seek to learn the underlying physics of electron–electron interactions directly.[6] As such, these approaches provide an additional layer of physical information beyond MLIPs by predicting quantum-chemical intermediates and electronic structure parameters, thereby enabling the acceleration of conventional electronic structure methods.[6] However, the vast majority of ML-based models developed to date adopt energy-related quantities, most commonly the total energy, total correlation energy, or closely related terms, as their primary training targets.[8-18] Other approaches include models that predict electronic-structure matrices,[19, 20] most notably the Hamiltonian (and overlap) matrix, as supervised outputs, such as the SchNOrb deep-learning framework,[21, 22] quantum Monte Carlo-based models,[23, 24] models that exploit learned functional relationships between dominant and enslaved coupled-cluster amplitudes to reduce the cost of coupled-cluster calculations,[25, 26] and electronic structure approaches that explicitly target one-electron reduced density matrices.[27]

In this context, we have developed a data-driven coupled-cluster scheme (DDCC) that exemplifies this integration of ML into wavefunction theory.[28-34] DDCC utilizes data from low-level correlated methods such as mean-field HF and second-order Møller–Plesset perturbation theory (MP2), to predict parameters of the higher-level CCSD or CCSD(T) wave functions. The predicted parameters can either be employed directly to approximate coupled-cluster correlation energies or utilized as improved initial guesses for the iterative CCSD solver.[30] In the former case, CCSD-quality correlation energies can be achieved at an effective MP2-level computational cost, while in the latter, the number of iterations required for convergence is significantly reduced. Earlier variants of the DDCC methodology relied on random-forest (RF) models, which indeed delivered remarkable accuracy at reduced computational cost.[32] Despite these advances, RF-based DDCC methods were constrained from two fundamental limitations. First, DDCC(RF) models for singles and doubles excitations exhibited poor portability due to the large memory footprints required to achieve high accuracy. Second, RF is not an optimum choice for handling large number of data, given the exponential growth of the two-electron excitations. Lastly, RF models inherently lack an efficient and scalable framework for multitask learning, which is essential for the simultaneous prediction of high-dimensional output quantities.

Based on these considerations and motivated by the versatility of neural networks to handle large datasets while incorporating physical constraints into complex architectures,[35] we introduce here the data-driven coupled-cluster neural networks (DDCCNet), a family of deep learning architectures designed to predict the coupled-cluster wave function. The DDCCNet models represent a substantial methodological advance over previous machine learning approaches that aim to capture electron correlation terms. The framework introduces native multitasking capabilities, a fully reengineered input vectorization of the one- and two-electron excitation processes, and physics-enhanced neural network architectures specifically tailored to the structure of the coupled-cluster solver. These developments are further reinforced by newly designed loss functions that enable more effective learning and significantly enhanced predictive accuracy. Collectively, these innovations allow the treatment of substantially larger molecular systems, well beyond the practical limits of earlier DDCC variants.

In this work, we present three versions of DDCCNet, each featuring a tailored multitask loss function designed to balance learning across the different prediction tasks. DDCCNet_v1 is a base platform required for testing and comparison with the more sophisticated versions 2 and 3 (DDCCNet_v2 and DDCCNet_v3, respectively). We demonstrate their versatility on the prediction of $CO_2$ cluster energies and a diverse dataset of small organic molecules, highlighting both their accuracy and transferability across chemically distinct regimes. Comparison of the DDCCNet models against the earlier DDCC based on RF on methanol conformers demonstrates a substantial improvement in energy prediction accuracy, underscoring the superiority of neural networks over ensemble tree methods for this task.

**Theoretical Aspects**

A brief overview of the coupled-cluster equations, along with the working expressions implemented in most quantum-chemical software packages, is provided in this section. Because these equations directly or indirectly introduced in our DDCCNet architectures presented here, their inclusion is essential for completeness. We have followed the same notation as in the work of Stanton et al..[36]

A direct variational solution of the coupled-cluster equations with respect to the $t_1$ and $t_2$ amplitudes is not nontrivial. Instead, these amplitudes are determined by solving the nonlinear projected equations obtained from the stationarity of the energy with respect to the corresponding excitation operators:

$$\langle\Psi_0|\exp(-\hat{T})\hat{H}\exp(\hat{T})|\Psi_0\rangle = E_{CC} \tag{1}$$

$$\langle\mu|\exp(-\hat{T})\hat{H}\exp(\hat{T})|\Psi_0\rangle = 0 \tag{2}$$

To improve both computational efficiency and memory usage, the residual CC $t_1$ and $t_2$ amplitudes are not evaluated directly, but expressed in terms of a series of intermediates, as shown in Eq. (3) and (4).[36] This strategy avoids redundant tensor contractions, reduces overall scaling, and minimizes storage demands, thereby making the coupled-cluster iterations more tractable for larger molecular systems.

$$\Delta t_i^a D_i^a = f_{ia} + \sum_e t_i^e F_{ae} - \sum_m t_m^a F_{mi} + \sum_{me} t_{im}^{ae} F_{me} - \sum_{nf} t_n^f \langle na\|if\rangle - \frac{1}{2}\sum_{mef} t_{im}^{ef}\langle ma\|ef\rangle \\ - \frac{1}{2}\sum_{men} t_{mn}^{ae} \langle nm\|ei\rangle \tag{3}$$

$$\Delta t_{ij}^{ab} D_{ij}^{ab} = \langle ij\|ab\rangle + P_-(ab)\sum_e t_{ij}^{ae}\left(F_{be} - \frac{1}{2}\sum_e t_m^b F_{me}\right) - P_-(ij)\sum_m t_{im}^{ab}\left(F_{mj} + \frac{1}{2} t_j^e F_{me}\right) \\ + \frac{1}{2}\sum_{mn} \tau_{mn}^{ab} W_{mnij} + \frac{1}{2}\sum_{ef} \tau_{ij}^{ef} W_{abef} \\ + P_-(ij)P_-(ab)\sum_{me}\left(t_{im}^{ae} W_{mbej} - t_i^e t_m^a \langle mb\|ej\rangle\right) + P_-(ij)\sum_e t_i^e \langle ab\|ej\rangle \\ - P_-(ab)\sum_m t_m^a \langle mb\|ij\rangle \tag{4}$$

In the above equations, $\Delta t_i^a$ is the residual term for $t_1$ amplitudes and $\Delta t_{ij}^{ab}$ is the residual term for $t_2$ amplitudes. The terms $D_i^a$ and $D_{ij}^{ab}$ are the energy differences between occupied and virtual orbitals:

$$D_i^a = \varepsilon_i - \varepsilon_a \tag{5}$$

$$D_{ij}^{ab} = \varepsilon_i + \varepsilon_j - \varepsilon_a - \varepsilon_b, \tag{6}$$

where $\varepsilon_p$ is the energy of the $p^{\text{th}}$ spin orbital.

For the calculation of residual $t_1$ and $t_2$ amplitudes, we need to derive a set of intermediates. The expressions for those intermediates are given below.

$$F_{ae} = (1 - \delta_{ae})f_{ae} - \frac{1}{2}\sum_m f_{me}t_m^a + \sum_{mf} t_m^f \langle ma \| fe \rangle - \frac{1}{2}\sum_{mnf} \tilde{\tau}_{mn}^{af} \langle mn \| ef \rangle \tag{7}$$

$$F_{mi} = (1 - \delta_{mi})f_{mi} + \frac{1}{2}\sum_e f_{me}t_i^e + \sum_{en} t_n^e \langle mn \| ie \rangle + \frac{1}{2}\sum_{nef} \tilde{\tau}_{in}^{ef} \langle mn \| ef \rangle \tag{8}$$

$$F_{me} = f_{me} + \sum_{nf} t_n^f \langle mn \| ef \rangle \tag{9}$$

$$W_{mnij} = \langle mn \| ij \rangle + P_-(ij) \sum_e t_j^e \langle mn \| ie \rangle + \frac{1}{4}\sum_{ef} \tau_{ij}^{ef} \langle mn \| ef \rangle \tag{10}$$

$$W_{mbej} = \langle mb \| ej \rangle + \sum_f t_j^f \langle mb \| ef \rangle - \sum_n t_n^b \langle mn \| ej \rangle - \sum_{nf} \left(\frac{1}{2}t_{jn}^{fb} + t_j^f t_n^b\right) \langle mn \| ef \rangle \tag{11}$$

$$Z_{mbij} = \sum_{ef} \tau_{ij}^{ef} \langle mb \| ef \rangle \tag{12}$$

$$\tau_{ij}^{ab} = t_{ij}^{ab} + t_i^a t_j^b - t_i^b t_j^a \tag{13}$$

$$\tilde{\tau}_{ij}^{ab} = t_{ij}^{ab} + \frac{1}{2}(t_i^a t_j^b - t_i^b t_j^a) \tag{14}$$

The permutation operator $P_-(ij)$ used above is defined as,

$$P_\pm(pq) = 1 \pm \mathcal{P}(pq) \tag{15}$$

where $\mathcal{P}(pq)$ permutates spin orbitals $p$ and $q$.

Additionally, there is another intermediate calculation during the process of determining $\Delta t_{ij}^{ab}$ that arises due to the spin-factorization of $W_{mbej}$:

$$W_{mbej} = -\langle mb\|je\rangle + \sum_f t_j^f \langle mb\|fe\rangle - \sum_n t_n^b \langle mn\|je\rangle - \sum_{nf}\left(\frac{1}{2}t_{jn}^{fb} + t_j^f t_n^b\right)\langle mn\|fe\rangle \quad (16)$$

**Model Development and Implementation**

In this work, we have developed three different variants of deep network architectures to predict simultaneously CCSD $t_1$ and $t_2$ amplitudes based on electronic structure data collected from HF and MP2.

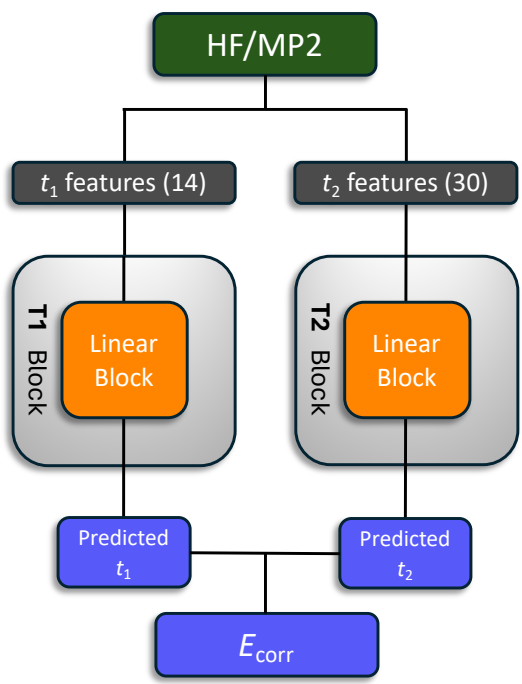

**Figure 1.** Architecture of DDCCNet_v1. Data are collected from HF and MP2 calculations (green box) and transformed into two input vectors (gray boxes), one related to the one-electron excitations ($t_1$ amplitudes) and a second related to the two-electron excitations ($t_2$ amplitudes). Orange boxes correspond to the linear layer of the neural network, blue boxes denote tasks related

to the loss function. The other two DDCCNet variants follow the same general architecture, with a more complex, physics-enhanced **T1** and **T2** blocks.

***DDCCNet_v1.*** The first DDCCNet architecture (DDCCNet_v1) consists of two distinct linear sub-networks (blocks) dedicated to predicting $t_1$ and $t_2$ amplitudes (Figure 1). We will refer to these as **T1** and **T2** blocks, respectively. The training data were constructed from calculations at the HF and MP2 levels, which provide the necessary information for both excitation manifolds. For the $t_1$ amplitude prediction (**T1** block), we compiled an input feature vector containing 14 descriptors that characterize the one-electron excitation processes in the coupled-cluster singles manifold (see Supporting Information Section S1). As input to the **T2** block, we adopted the original thirty-dimensional feature representation previously introduced in DDCC,[28] which encodes the relevant two-electron excitation information. Each linear block consists of seven fully connected layers, with 196 neurons per hidden layer and ReLU activation functions introducing nonlinearity between layers. The parameters of both sub-networks are optimized jointly using a composite loss function that enforces the physical interdependence between the $t_1$ and $t_2$ amplitudes (*vide infra*).

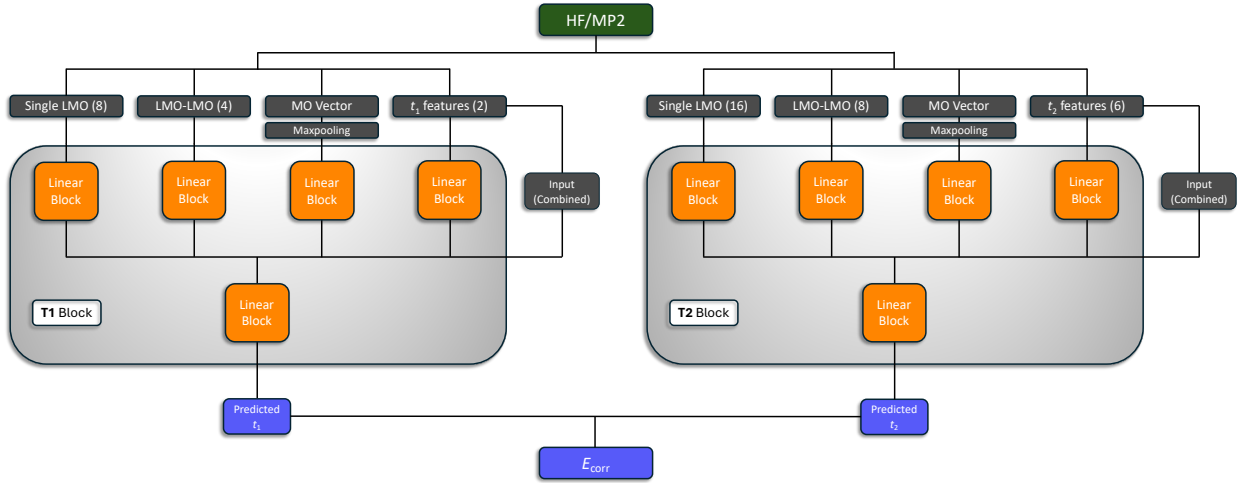

**Figure 2.** Architecture of DDCCNet_v2. Data are collected from HF and MP2 calculations (green box) and transformed into eight input vectors (gray boxes), four related to the one-electron excitations ($t_1$ amplitudes) and four related to the two-electron excitations ($t_2$ amplitudes), before

they are introduced into individual linear blocks (orange boxes). A combined vector containing all $t_1$ and $t_2$ input features is provided to a final linear block, together with the concatenated outputs of the individual blocks. Similar to DDCCNet_v1, one- and two-electron amplitudes are treated differently withing the **T1** and **T2** blocks, respectively. Blue boxes denote tasks related to the loss function.

***DDCCNet_v2.*** In the second version of DDCCNet (DDCCNet_v2), the feature vector is partitioned into four sections, and four different linear networks are used to predict $t_1$ and $t_2$ amplitudes, as shown in Figure 2. The four input sections are:

- **Single localized orbital features ($h_{LMO}$):** Orbital energies, one-electron Hamiltonian term, Coulomb and exchange contribution to orbital energies for a total of 8 and 16 features for the **T1** and **T2** blocks, respectively.
- **Localized orbital pair features ($h_{LMO-LMO}$):** Coulomb and exchange integral between occupied and virtual orbitals and log values of their magnitudes for a total of 4 and 8 features for the **T1** and **T2** blocks, respectively.
- **LMO vector ($h_{LMOvec}$):** Atomic orbital coefficients from Boys LMOs. A max-pooling layer is employed to aggregate variable-length features into a fixed-dimensional input representation for the model.
- ***$t_1$/$t_2$ amplitude features ($h_{t_1/t_2}$):*** DDCCNet_v2 employs a reduced feature set relative to DDCCNet_v1: 2 of the 14 original features are retained in the **T1** block, and 6 out of the original 30 features in the **T2** block (for more details, see Supporting Information Section S2).

Features of each separate section are introduced in four separate blocks of linear layers. Each of the four linear layer blocks contains seven linear layers with 196 nodes and ReLU activation functions in each hidden layer. Then the output from each linear block is concatenated and passed through a combined block (Eq. 17) with architecture identical to the previous linear block (seven linear layers with 196 nodes and ReLU activation functions in each hidden layer):

$$h = h_{\text{LMO}} \oplus h_{\text{LMO}-\text{LMO}} \oplus h_{\text{LMOvec}} \oplus h_{t_1/t_2} \qquad (17)$$

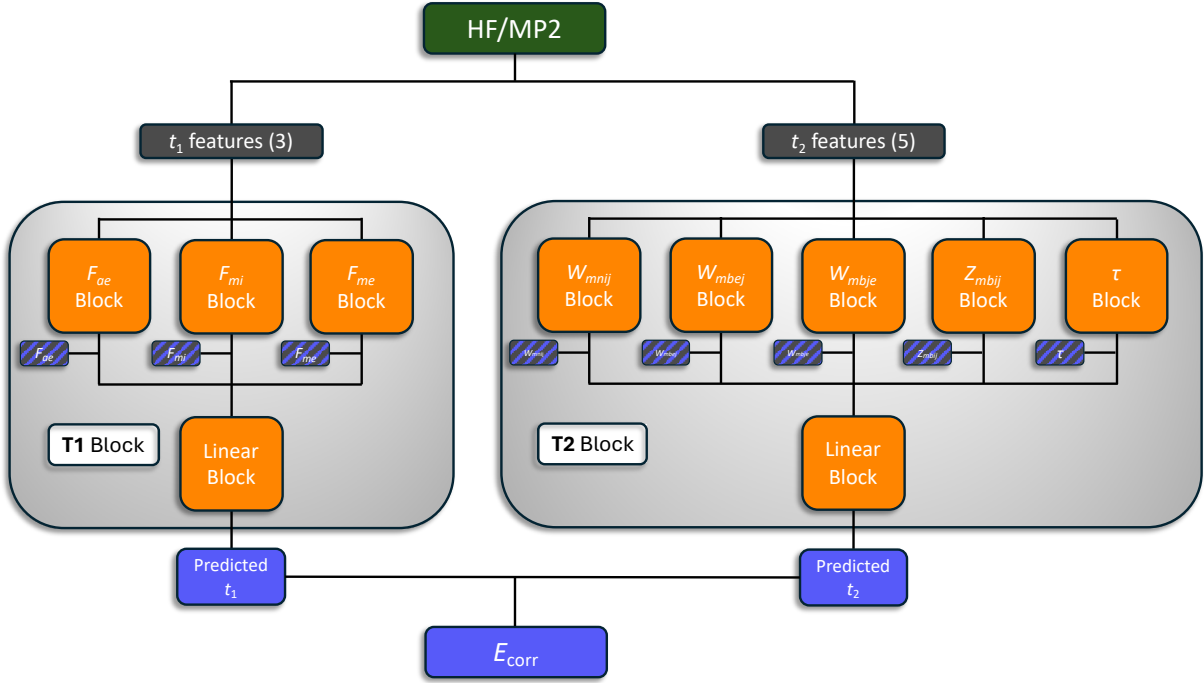

**Figure 3.** Architecture of DDCCNet_v3. Data are collected from HF and MP2 calculations (green box) and transformed into two vectors (gray boxes), one as input to the **T1** block (three features) and one as input to the **T2** block (five features). Blue and blue/black boxes denote tasks related to the loss function. The blue/black boxes are the individual loss functions per indermediate.

*DDCCNet_v3.* The third version of DDCCNet (DDCCNet_v3) aims at implementing the actual coupled-cluster working equations [Eqs. (7) – (16)] directly into the network architecture to achieve an enhanced physics-informed learning framework. The complete architecture of DDCCNet_v3 is shown in Figure 3. The **T1** block consists of three separate blocks which are updated via the intermediates $F_{ae}$, $F_{mi}$ and $F_{me}$ described in Eqs. (7) – (9), respectively. The final layer from these three blocks in concatenated and passed through a linear block:

$$h = h_{ae} \oplus h_{mi} \oplus h_{me} \tag{18}$$

where $h_{ae}$, $h_{mi}$, and $h_{me}$ correspond to the four layers from the $F_{ae}$, $F_{mi}$ and $F_{me}$ blocks, respectively.

Similarly, **T2** block has five separate blocks of linear layers for the five intermediates, $W_{mnij}$, $W_{mbej}$, $W_{mbje}$, $Z_{mbij}$ and $\tau$ described in Eqs. (10)-(13), respectively. The final layer of these intermediate blocks is concatenated and passed through a block of linear layers to predict $t_2$ amplitudes.

$$h = h_{mnij} \oplus h_{mbej} \oplus h_{mbje} \oplus h_{mbij} \oplus h_{\tau} \tag{19}$$

Multiple linear blocks that are identical to those of DDCCNet_v1 and DDCCNet_v2 (each linear block contains seven linear layers with 196 nodes and ReLU activation functions in each hidden layer) are used for two separate parts of the DDCCNet_v3 model as well (i.e. the **T1** and **T2** blocks). A reduced dimension input vector that included the most important features with respect to the targeted amplitudes was used (see Supporting Information Section S3).

***Loss Functions.*** DDCCNet_v1 and DDCCNet_v2, a combination of three loss functions was used. The following metrics were considered in the loss functions:

$$\text{MSE}_{t_1} = \frac{\sum(t_1 - t_1')^2}{n_{t_1}} \tag{17}$$

$$\text{MSE}_{t_2} = \frac{\sum(t_2 - t_2')^2}{n_{t_2}} \tag{18}$$

$$\text{MAE}_{t_1} = \frac{\sum |t_1 - t_1'|}{n_{t_1}} \qquad (19)$$

$$\text{MAE}_{t_2} = \frac{\sum |t_2 - t_2'|}{n_{t_2}} \qquad (20)$$

$$\text{MAE}_{\text{corr}} = |E_{\text{corr}} - E_{\text{corr}}'| \qquad (21)$$

$$\text{RSS} = \sqrt{\sum (t_1 - t_1')^2 + \sum (t_2 - t_2')^2} \qquad (22)$$

MSE is the mean signed error [Eqs. (17) and (18)], MAE is the mean average error [Eqs. (19) - (21)], and RSS is the residual sum of squares [Eq. (22)] Here, $t_1'$ and $t_2'$ are the predicted $t_1$ and $t_2$ amplitudes, respectively, and $t_1$ and $t_2$ are the exact CCSD one and two-electron amplitudes. The total number of $t_1$ and $t_2$ amplitudes in a batch are denoted by $n_{t_1}$ and $n_{t_2}$ respectively.

For DDCCNet_v1, we use the following loss functions for the **T1** and **T2** blocks:

$$\text{Total Loss}_{t_1}^{v1} = \text{MSE}_{t_1} + \text{RSS} + \text{MAE}_{\text{corr}} \qquad (23)$$

$$\text{Total Loss}_{t_2}^{v1} = \text{MSE}_{t_2} + \text{RSS} + \text{MAE}_{\text{corr}} \qquad (24)$$

In Eq. (23) and (24), the superscript v1 indicates that these expressions of the total loss correspond to the DDCCNet_v1 model. For the DDCCNet_v2 models, systematic model optimization showed that the most effective loss functions include $\text{MAE}_{t_1}$ and $\text{MAE}_{t_2}$ instead of $\text{MSE}_{t_1}$ and $\text{MSE}_{t_2}$, respectively. Therefore, the total loss functions for $t_1$ and $t_2$ are:

$$\text{Total Loss}_{t_1}^{v2} = \text{MAE}_{t_1} + \text{RSS} + \text{MAE}_{\text{corr}} \qquad (25)$$

$$\text{Total Loss}_{t_2}^{v2} = \text{MAE}_{t_2} + \text{RSS} + \text{MAE}_{\text{corr}} \qquad (26)$$

As mentioned above, for DDCCNet_v3, we use blocks of linear layers to predict each intermediate; three for **T1** block [Eqs. (7) – (9)] and five for **T2** block [Eqs. (10) – (16)]. Additional loss functions were added to the **T1** and **T2** blocks:

$$\text{MSE}_I = \frac{\sum (I - I')^2}{n_I} \qquad (27)$$

Here $I$ denotes the value of the intermediate and $I'$ denotes the predicted value of the same intermediate. The number of intermediates per batch is denoted by $n_I$. The total loss function for **T1** and **T2** blocks are:

$$\text{Total Loss}_{t_1}^{v_3} = \text{MSE}_{t_1} + \text{RSS} + \text{MAE}_{\text{corr}} + \sum_{I=1}^{3} \text{MSE}_{I,t_1} \qquad (28)$$

$$\text{Total Loss}_{t_2}^{v_3} = \text{MSE}_{t_2} + \text{RSS} + \text{MAE}_{\text{corr}} + \sum_{I=1}^{5} \text{MSE}_{I,t_2} \qquad (29)$$

*Amplitude Space Reduction:* Due to the large volume and uneven distribution of amplitudes generated for training and testing, the initial DDCCSD models encountered two primary challenges. The first was the computational difficulty associated with managing, training, and storing machine learning models on such extensive datasets. The second challenge was the tendency of the models to overfit, primarily caused by the disproportionate number of amplitudes clustered near zero. To address these issues, we employ the large amplitude (LA) scheme, an amplitude sampling scheme previously developed for the DDCC(RF) models.[32] The LA approach retains only MP2 amplitudes whose magnitudes exceed a predefined cutoff, thereby excluding small amplitudes from DDCC and DDCCNet training. In our previous work, we demonstrated that a cutoff of $1\times10^{-4}$ yields consistently accurate results across a broad range of chemical systems. Accordingly, the same threshold is adopted for DDCCNet.

In the development of the three DDCCNet variants, we also took under consideration the symmetry of the amplitudes:

$$t_{ij}^{ab} = t_{ji}^{ba} \qquad (30)$$

The compact $t_2$ vector with the predicted amplitudes is then unpacked by using the same symmetry consideration of Eq. (30), and the final CCSD correlation energies are computed.

**Computational Details**

Neural network models were implemented within the PyTorch deep learning framework.[37] All electronic structure calculations and data extraction were performed with the Psi4NumPy quantum chemistry package.[38, 39] The cc-pVDZ basis set[40] together with Boys localized orbitals[41] were used in all CCSD calculations.

Methanol conformers were used to train and test the models. A total of 50 conformers were generated for training by varying the O–H bond distance between 0.80–1.00 Å and the C–O–H bond angle between 105.5–110.0°. For testing, 25 conformers were constructed with O–H bond distances between 0.85–0.95 Å and C–O–H bond angles between 105.5–109.5°. For comparison purposes, the same molecular sets were used for training and testing the DDCC(RF) models.[32]

The second application focused on the calculation of the CCSD energies of $CO_2$ clusters. Training was performed using $CO_2$ monomers, dimers, and trimers. Monomer geometries were generated by varying the C–O bond length from 1.16–1.19 Å and the O–C–O bond angle from 175.0–180.0°. Dimer and trimer structures were obtained from LAMMPS[42] molecular dynamics simulations at 300 K employing the GROMOS_54A7 force field.[43, 44] For testing, ten dimer structures, five trimer structures, three tetramer structures, and three pentamer structures were generated using the same simulation protocol as for the dimers and trimers.

For the third application, a diverse set of organic molecules was obtained from the GDB11 database.[45] A total of 275 molecules containing five non-hydrogen atoms (C, N, and O) were selected. Hereby this molecular database will be referred as GDB5′ database. The SMILES strings of the GDB5′ molecules were converted to three-dimensional geometries using the RDKit package[46] with the Experimental-Torsion-Knowledge Distance Geometry (ETKDG v3) method.[47] All DDCCNet models reported in the next section were trained with 200 randomly selected molecules and tested on the remaining 75 molecules.

**Results and Discussion**

*Methanol Conformers:* The performance of different DDCC variants for predicting methanol conformer energies is summarized in Table 1. The DDCC model employing a Random Forest algorithm, denoted as DDCC(RF), was used as a baseline for comparison with the neural-network architectures. This ensemble-based approach, which relies on decision-tree averaging rather than gradient-based optimization, yielded a mean absolute error (MAE) of 7.629 m$E_h$. The resulting performance is more than an order of magnitude less accurate than that achieved with the DDCCNet variants, highlighting the superior representational capability of the latter in capturing the nonlinear relationships governing coupled-cluster amplitudes.

**Table 1.** Comparison of mean absolute errors (MAE) for CCSD errors in predicting methanol molecules using the DDCC(RF) and DDCCNet models. All energies in m$E_h$.

| Model | Test MAE (m$E_h$) | Train $R^2$ | Test $R^2$ | Maximum Test Error (m$E_h$) |
|---|---|---|---|---|
| DDCC(RF) | 5.894 | 0.9125 | 0.8662 | 8.711 |
| DDCCNet_v1 | 0.251 | 0.9902 | 0.9616 | 0.642 |
| DDCCNet_v2 | 0.229 | 0.9961 | 0.9886 | 0.473 |
| DDCCNet_v3 | 0.198 | 0.9939 | 0.9832 | 0.626 |

In contrast, the DDCCNet series of models demonstrated excellent predictive accuracy. The first version, DDCCNet_v1 achieved an MAE of 0.251 m$E_h$, already significantly outperforming the DDCC(RF) model. Subsequent architectural refinements in DDCCNet_v2 and DDCCNet_v3 led to further improvements, reducing the MAE to 0.229 m$E_h$ and 0.198 m$E_h$, respectively. The progressive reduction in error across model versions suggests that the modifications enhanced the network's ability to capture the complicated correlations between low-level electronic structure features and CCSD wave function parameters and electronic energies. Importantly, all three neural network models achieve sub-milli Hartree accuracy and within sub-chemical accuracy (~0.1 kcal/mol or 0.5 kJ/mol). The improvements from v1 to v3 also indicate that the DDCCNet framework is robust and scalable, and further optimization may yield even lower errors. These results establish the feasibility of employing deep learning approaches, specifically DDCCNet architectures, to replace or accelerate electronic structure calculations for methanol and potentially more complex molecular systems.

*$CO_2$ Conformers:* The performance of DDCCNet models for $CO_2$ clusters of increasing size is summarized in Table 2. The DDCCNet variants were trained on monomers, dimers, and trimers, and subsequently tested on dimers through pentamers.

**Table 2.** Comparison of mean absolute errors (MAEs) for CCSD energy calculations of $CO_2$ clusters. All errors in m$E_h$.

| $CO_2$ Test Set | DDCCSD_v1 | | | DDCCSD_v2 | | | DDCCSD_v3 | | |
|---|---|---|---|---|---|---|---|---|---|
| | MAE | MAE per atom | MAE per electron | MAE | MAE per atom | MAE per electron | MAE | MAE per atom | MAE per electron |
| Dimers | 3.402 | 0.567 | 0.077 | 0.913 | 0.152 | 0.021 | 1.386 | 0.231 | 0.032 |
| Trimers | 6.887 | 0.765 | 0.104 | 0.812 | 0.090 | 0.012 | 1.062 | 0.118 | 0.016 |
| Tetramers | 15.752 | 1.313 | 0.179 | 0.923 | 0.077 | 0.010 | 4.191 | 0.349 | 0.048 |
| Pentamers | 17.088 | 1.139 | 0.155 | 1.000 | 0.067 | 0.009 | 6.578 | 0.439 | 0.060 |

The performance of DDCCNet_v1 is unsatisfactory. For $CO_2$ dimers, a MAE of 3.402 m$E_h$ indicates insufficient predictive accuracy. This deficiency becomes more pronounced for larger $CO_2$ clusters, with the MAEs increasing up to 17.088 m$E_h$ for $CO_2$ pentamers (1.139 m$E_h$ per atom or 0.155 per electron). On the contrary, the DDCCNet_v2 model demonstrated high accuracy which is consistent and transferable across the different test sets. The dimer prediction yielded an overall MAE of 0.913 m$E_h$ (0.152 m$E_h$ per atom or 0.012 m$E_h$ per electron), while the trimer prediction improved to 0.812 m$E_h$ (0.090 m$E_h$ per atom). Notably, even when extrapolating to larger systems outside the training domain, such as $CO_2$ tetramers and pentamers, the errors remained modest at 0.923 m$E_h$ (0.077 m$E_h$ per atom or 0.010 m$E_h$ per electron) and 1.000 m$E_h$ (0.067 m$E_h$ per atom or 0.009 m$E_h$ per electron), respectively. The decreasing MAE per atom with cluster size indicates that the model generalizes well. These results suggest that DDCCNet_v2 captures the essential many-body interactions governing cluster energetics. In contrast, DDCCNet_v3 displayed lower accuracy and poorer transferability beyond the training set. While the dimer and trimer results remained reasonable with MAEs of 1.386 m$E_h$ (0.231 m$E_h$ per atom) and 1.062 m$E_h$ (0.118 m$E_h$ per atom), respectively, the errors increased sharply for larger clusters. For tetramers, the MAE increased to 4.191 m$E_h$ (0.349 m$E_h$ per atom), and for pentamers it increased further to 6.578 m$E_h$ (0.439 m$E_h$ per atom). This systematic decline in performance indicates that modifications in DDCCNet_v3, while beneficial for the methanol system, compromised the ability of the model to generalize across $CO_2$ cluster sizes. The rapid error growth suggests difficulties in capturing long-range and higher-order many-body correlations, which become increasingly significant in larger clusters.

Overall, these results highlight a trade-off between model architecture improvements for small molecules (as evidenced by methanol performance) and transferability to larger systems (as seen in $CO_2$ clusters). DDCCNet_v2 appears more robust in extrapolation, maintaining chemically accurate predictions across monomers to pentamers. Meanwhile, DDCCNet_v3 requires further refinement to recover transferability while preserving the accuracy gains observed for smaller systems.

***Small Organic Molecules.*** The performance of DDCCNet_v2 was further evaluated on a diverse set of 275 organic molecules containing five non-hydrogen atoms (C, N, and O) extracted from the GDB11 database. the following discussion, this dataset is referred to as GDB5′. From the full dataset of 275 molecules, 75 were held out as a fixed test set. The remaining molecules were used to construct multiple training subsets of increasing size (from 10 up to 200) to assess how model performance scales with training set size. Figure 4 shows the variation of the MAE of the total correlation energy as a function of the number of molecules included in the training set.

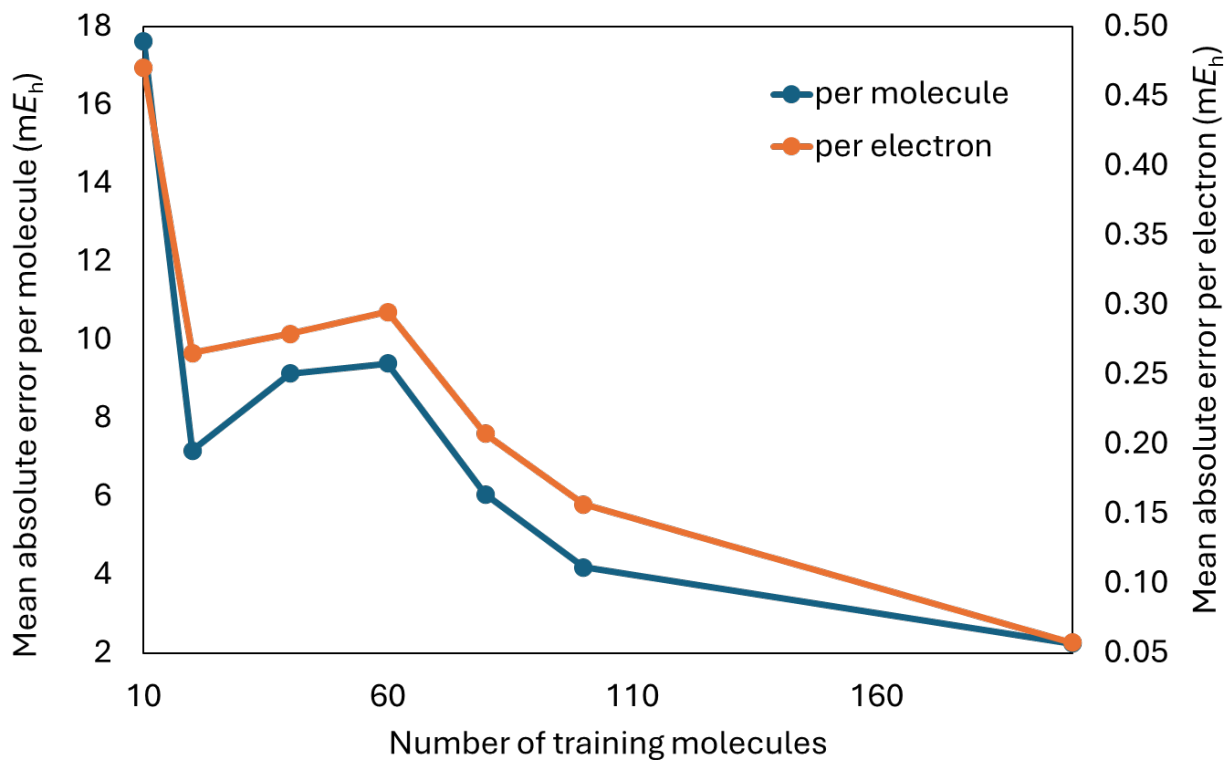

**Figure 4.** Mean absolute error in m$E_h$ per molecule (in blue) and per electron (in orange) predicted by DDCCNet_v2 as a function of the number of training molecules in GDB5′. Left y-axis corresponds to the MAE per molecule, and right axis corresponds to the MAE per electron.

As shown in Figure 4, the MAE decreases systematically with increasing training set size, indicating that the model effectively captures the underlying nonlinear relationships between molecular orbital features and electron correlation energy. A systematic increase in accuracy is observed once the model is trained with data from more than 10 molecules (MAE of 17.625 m$E_h$). When the full dataset was used (200 training molecules) the error dropped to 2.245 m$E_h$. This trend demonstrates that DDCCNet_v2 efficiently learns transferable features from limited data and reaches a near-asymptotic accuracy with moderate dataset sizes. Training with larger molecular datasets are expected to reduce these errors around 1.0 m$E_h$.

The corresponding per-atom and per-electron MAEs follow a similar trend, reaching 0.449 m$E_h$ per atom and 0.057 m$E_h$ per electron for the models trained on the dataset of 200 molecules. The decreasing standard deviation (from 13.5 to below 1.8 m$E_h$) further confirms the stability and consistency of the model predictions.

The smooth convergence behavior, combined with reduced variance, highlights the efficiency of the multitask learning approach and the amplitude sampling selection strategy that is introduced in the DDCCNet, which prioritizes physically significant amplitude regions while reducing noise from low-magnitude, near-zero amplitudes. These findings confirm that DDCCNet_v2 generalizes well to unseen molecular structures and provides chemically accurate predictions for larger molecular systems at significantly reduced computational cost compared to full CCSD calculations.

**Conclusions**

In this work, we presented the development and comprehensive evaluation of the data-driven coupled-cluster neural network (DDCCNet) framework, designed to predict CCSD amplitudes and correlation energies through multitask, physics-informed deep learning. Three physics-enhanced deep neural networks of increasing sophistication were developed, all utilizing data from HF and MP2 calculations and sharing the same overall multi-task architecture. The first one (DDCCNet_v1) consists of two neural network branches: one dedicated to predicting the $t_1$ amplitudes and the other to the $t_2$ amplitudes. DDCCNet_v2 extends the first model by introducing additional one- and two-electron information that describe the $t_1$ and $t_2$ amplitudes, respectively. The third DDCCNet variant (DDCCNet_v3) incorporates additional information inspired by the projected coupled-cluster equations and is trained using data from coupled-cluster intermediates that are routinely computed and stored at each iteration. Model optimization considered the number and type of features derived from HF and MP2 calculations, the number and placement of hidden layers within each architecture, and the choice of loss function. Additional considerations were incorporated regarding amplitude selection for model training and the enforcement of amplitude symmetries.

The performance of the three DDCCNet models was first evaluated using methanol energies. All models were trained on multiple methanol conformations and achieved high accuracy, with MAEs below 0.25 m$E_h$ for all three DDCCNet variants. We next assessed the transferability of the DDCCNet models by training on small $CO_2$ clusters and predicting energies for larger aggregates. Specifically, the models were trained using data from $CO_2$ monomers, dimers, and trimers, and were evaluated on dimers and trimers as well as on larger, out-of-distribution $CO_2$ clusters comprising tetramers and pentamers. DDCCNet_v1 exhibited large deviations, reaching up to 17 m$E_h$, rendering it unsuitable for practical applications. DDCCNet_v3 achieved reasonable accuracy for dimers and trimers (MAEs of approximately 1 m$E_h$), but showed limited extrapolation capability for larger systems, with MAEs of 4.191 and 6.578 m$E_h$ for tetramers and pentamers, respectively. In contrast, DDCCNet_v2 demonstrated consistent performance across both in-distribution and out-of-distribution cluster sizes. For $CO_2$ pentamers, the MAE in the total correlation energy was 1.000 m$E_h$, corresponding to 0.067 m$E_h$ per atom and only 0.009 m$E_h$ per electron. Finally, we assessed the applicability of DDCCNet_v2 to small organic molecules of diverse chemical composition. Models were trained using between 10 and 200 molecules and evaluated on a fixed set of 75 out-of-distribution organic molecules, yielding predictive uncertainties of approximately 2 m$E_h$. The resulting learning curve exhibits a systematic reduction in MAE with increasing training set size, highlighting efficient learning behavior and robust generalization. Collectively, these results indicate that DDCCNet_v2 captures the nontrivial relationships between molecular orbital features and correlated-electron amplitudes with high accuracy across chemically diverse molecular systems.

Overall, DDCCNet demonstrates the advantages of integrating machine learning with wavefunction-based quantum chemistry. By embedding physical insight directly into the network architecture and training strategy, DDCCNet provides a transferable, data-efficient, and systematically improvable framework for high-accuracy electronic structure prediction. Relative to traditional regression approaches and earlier DDCCSD implementations, DDCCNet offers several key advantages, including scalability through neural-network architectures capable of handling large and heterogeneous datasets, improved physical consistency via multitask feature partitioning and amplitude symmetries, and efficient convergence enabled by adaptive learning optimization. Collectively, these features allow DDCCNet to model correlation energies with high accuracy at a fraction of the computational cost of conventional CCSD methods. Ongoing developments focus on reducing remaining computational overheads and extending the framework to larger and more chemically diverse systems, with the goal of enabling general-purpose, physics-informed machine learning models for electronic structure across a broad range of chemical domains.

## Acknowledgments

The authors gratefully acknowledge the CAREER and CAS-Climate programs of the National Science Foundation for financial support of this work (Grant no. CHE-2143354). The authors acknowledge the Infrastructure for Scientific Applications and Advanced Computing (ISAAC) of the University of Tennessee for computational resources.

## Supporting Information Available

The code base containing the DDCCNet models can be found in the following GitHub repository: https://github.com/Varunasp93/DDCCNet_public